\documentclass[traditabstract]{aa}
\usepackage{txfonts}
\usepackage{graphicx}
\usepackage{natbib}

\begin{document}
\title{Cluster radius and sampling radius in the determination
       of cluster membership probabilities}
\author{N\'estor S\'anchez \and Bel\'en Vicente \and Emilio J. Alfaro}
\institute{Instituto de Astrof\'{\i}sica de Andaluc\'{\i}a,
           CSIC, Apdo. 3004, E-18080, Granada, Spain\\
           \email{nestor@iaa.es}}
\date{November 30, 2009}
\abstract{We
analyze the dependence of the membership probabilities
obtained from kinematical variables on the radius of the
field of view around open clusters (the sampling radius,
$R_s$). From simulated data, we show that the best
discrimination between cluster members and non-members
is obtained when the sampling radius is very close to
the cluster radius. At higher $R_s$ values more field
stars tend to be erroneously assigned as cluster members.
From real data of two open clusters (NGC~2323 and NGC~2311)
we obtain that the number of identified cluster members
always increases with increasing $R_s$. However, there is
a threshold $R_s$ value above which the identified cluster
members are severely contaminated by field stars and the
effectiveness of membership determination is relatively
small. This optimal sampling radius is $\simeq 14$ arcmin
for NGC~2323 and $\simeq 13$ arcmin for NGC~2311. We discuss
the reasons for such behavior and the relationship between
cluster radius and optimal sampling radius. We suggest that,
independently of the method used to estimate membership
probabilities, several tests using different sampling
radius should be performed in order to evaluate the
existence of possible biases.}
\titlerunning{Cluster radius and sampling radius in open clusters}
\authorrunning{S\'anchez et al.}
\maketitle

\section{Introduction}

The large astrometric catalogues derived from surveys covering
very wide areas of the sky are allowing the systematic searching
of new star systems \citep[see, for example,][and references
therein]{Lop98,Hoo99,Kaz02,Myu03,Cab08b,Zha09}. The searching
process is based on the detection of well defined structures in
some subsets of the phase space. The presence of both spatial density
peaks and proper motion peaks indicates the existence of star clusters;
peaks visible only in the proper motion distributions suggest the
existence of moving groups; whereas more spread and less dense
velocity-position correlated structures could be associated to
stellar streams. Once these structures have been detected, the 
next step is to search for identify possible members of the star
system. For the particular case of open clusters, the most
often used procedure to select possible cluster members is the
algorithm designed by \citet{San71}. This algorithm is based on 
a former model proposed by \citet{Vas58} for the proper motion
distribution. The model assumes that cluster members and field
stars are distributed according to circular and elliptical
bivariate normal distributions, respectively.
The Sanders' algorithm, or some variation or refinement
of it, has been and still is widely used to estimate cluster
memberships either as the only method or as part of a more
complet treatment that includes, for example, spatial and/or
photometric criteria. Some recent representative references
are \citet{Wu02,Jil03,Bal04,Dia06,Kra07,Wir09}.

With the advent of large catalogues and databases available
via internet and future surveys such as the forthcoming
{\it Gaia} mission of ESA, the interest in developing and 
applying fully automated techniques is increasing among
the astronomical community. However, special care must be
taken to avoid obtaining biased results. In this work we
will show that the results yielded when using the Sanders'
algorithm significantly depend on the choice of the size
of the field of view surrounding the cluster. So, once
detected a possible open cluster, it is natural to ask
what area of the sky should be sampled in order to get
the most reliable membership determinations. It is
equally important to ask about the robustness of used
methodology, i.e. how does the solution change when
the sampled area is varied? Here we explore these
subjects by using both simulated and real data.
In Section~\ref{metodo} we briefly present the
method used to determine memberships and describe
the simulations that we performed to analyze the
expected behavior. The results of applying
the Sanders' algorithm on the simulated data
are discussed in Section~\ref{resultados}.
After this, in Section~\ref{cumulos} we use real astrometric
data of two open clusters (NGC~2323 and NGC~2311) to evaluate
the performance of the algorithm. We discuss strategies to
estimate the optimal sampling radius, i.e. the maximum radius
beyond which the identified cluster members are expected to be 
severely contaminated by field stars. The main results of the
present work are summarized in Section~\ref{conclusiones}.

\section{Description of the method}
\label{metodo}

\subsection{Membership determination}
\label{membresias}

The key point of the membership discrimination method
is the assumption that the distribution of observed
proper motions ($\mu_x$,$\mu_y$) can be described
by means of two bivariate normal distributions,
one circular for the cluster and one elliptical
for the field \citep{Vas58}. Let $\Phi_c$ and
$\Phi_f$ be the cluster and field probability
density functions, respectively. Then,
\begin{equation}
\Phi_c(\mu_x,\mu_y) = \frac{1}{2\pi\sigma_c^2}
\exp\left\{ -\frac{1}{2} \left[ \left(
\frac{\mu_x - \mu_{x,c}}{\sigma_c} \right)^2 +
\left( \frac{\mu_y - \mu_{y,c}}{\sigma_c} \right)^2
\right] \right\}
\end{equation}
and
\begin{eqnarray}
\Phi_f(\mu_x,\mu_y) = & \frac{1}{2\pi\sigma_{x,f}\sigma_{y,f}
\sqrt{1-\rho^2}}
\exp\left\{ -\frac{1}{2(1-\rho^2)} \left[ \left(
\frac{\mu_x - \mu_{x,f}}{\sigma_{x,f}} \right)^2 \right. \right.
\\ \ & + \left. \left.
\left( \frac{\mu_y - \mu_{y,f}}{\sigma_{y,f}} \right)^2
-2\rho\left(\frac{\mu_x-\mu_{x,f}}{\sigma_{x,f}}\right)
\left(\frac{\mu_y-\mu_{y,f}}{\sigma_{y,f}}\right)
\right] \right\}\ \ ,
\nonumber
\end{eqnarray}
where $(\mu_{x,c},\mu_{y,c})$ is the cluster
distribution centroid with standard deviation
$\sigma_c$, $(\mu_{x,f},\mu_{y,f})$ is the
field centroid with standard deviations
$\sigma_{x,f}$ and $\sigma_{y,f}$, and
$\rho$ is the correlation coefficient
of field stars. The probability density
function for the whole sample is simply
\begin{equation}
\Phi(\mu_x,\mu_y) = n_c \Phi_c(\mu_x,\mu_y) +
n_f \Phi_f(\mu_x,\mu_y)\ \ ,
\end{equation}
$n_c$ and $n_f$ being the normalized numbers
of cluster and field stars, respectively.
For obtaining the unknown parameters (centroids,
standard deviations, numbers of members and
non-members) an iterative procedure is used
by applying the maximum likelihood principle
\citep{San71}. Here we use the algorithm
proposed by
\citet{Cab85}, which first detects and removes
outliers that can produce unrealistic solutions,
and then uses a more robust and efficient
iterative procedure for the model parameter
estimation.
Once these parameters are known, then membership
probability of the $i$-th stars can be calculated
directly as
\begin{equation}
p(i) = \frac{n_c \Phi_c(i)}{\Phi(i)}\ \ .
\end{equation}

\subsection{Simulations}
\label{simulaciones}

Let us consider a cluster with a given radius $R_c$. We are
defining ``cluster radius" as the radius of the smallest
circle that can completely enclose its stars. In real
situations $R_c$ is essentially an unknown quantity that
has to be estimated a posteriori, but here its value is
known and kept constant within each simulation. The total
number of stars belonging to the cluster is denoted by
$N_{c,max}$ and the number of field stars lying exactly
within the same sky area of the cluster is $N_{f,cri}$.
The independent variable is the radius of the field
encircling the cluster. This radius might represent
the radius of the field in which the observations are
made or the field around the cluster extracted from
an astrometric catalogue. We call this variable
the sampling radius $R_s$, which can be larger or
smaller than the cluster radius $R_c$.

The numbers of cluster stars and field stars to
be simulated are represented by $N_{c,sim}$ and
$N_{f,sim}$, respectively. Obviosuly, the number
of the number of cluster stars and field stars
{\it within the field of view} depend on the
size of this field, that is, both $N_{c,sim}$
and $N_{f,sim}$ are functions of $R_s$. If the
field stars distribute nearly uniformly in space
then $N_{f,sim}$ should increase as the sampling
radius increases as
\begin{equation}
\label{nfield}
N_{f,sim} (R_s) = N_{f,cri} \left( R_s/R_c \right) ^2 \ \ .
\end{equation}
The rate at which $N_{c,sim}$ increases with $R_s$
depends instead on the radial profile of the
surface density of cluster stars ($\Sigma_{c,sim}$).
For simplicity, let us assume that the surface
density at $r$ is given by \citep{Cab08a}
\begin{equation}
\label{densup}
\Sigma_{c,sim} (r) = \frac{\delta N_{c,max}}{2\pi R_c^\delta}
r^{\delta - 2} \ \ ,
\end{equation}
with the index $\delta \leq 2$. For the extreme case $\delta = 2$,
we have $\Sigma_{c,sim} = N_{c,max} /(\pi R_c^2) = constant$.
Integrating equation~(\ref{densup}) we obtain
the number of cluster stars within a given sampling
radius (for $R_s \leq R_c$),
\begin{equation}
\label{ncluster}
N_{c,sim} (R_s) = N_{c,max} \left( R_s/R_c \right) ^\delta \ \ .
\end{equation}
Negative $\delta$ values make no sense, so this approach is
limited to the range $0 < \delta \leq 2$. The role of the
parameter $\delta$ is just to control how fast $N_{c,sim}$
increases as $R_s$ increases. Thus, the exact functional
form is not needed to be known as long as we are able to
simulate either completely flat ($\delta = 2$) or extremely
peaked ($\delta \simeq 0$) density profiles.

To perform the simulations we distribute $N_{f,sim}$
field stars and $N_{c,sim}$ cluster stars according
to bivariate gaussian distributions in the proper motion
space $(\mu_x,\mu_y)$. The routine ``gasdev" from the Numerical
Recipes package \citep{Pre92} is used for generating
normally distributed random numbers. The fields are centered
at $(0,0)$ with standard deviations of $\sigma_{x,f} =
\sigma_{y,f} = \sigma_f$. The tests performed using
elliptical (rather than circular) distributions for the
field stars yielded essentially the same results and
trends. The clusters are centered at $(\mu_{x,c},\mu_{y,c})$
and have standard deviations $\sigma_{x,c} = \sigma_{y,c} =
\sigma_c$. Thus, for a given sampling radius $R_s$ and
according to equation~(\ref{nfield}), we randomly generate
$N_{f,sim}$ field stars that follow a bivariate normal
distribution in the proper motion space. For the cluster,
we generate $N_{c,sim}$ stars according to 
equation~(\ref{ncluster}) when $R_s \leq R_c$
and we generate $N_{c,sim} = N_{c,max}=constant$
stars when $R_s \geq R_c$. The three free parameters,
excluding those describing the gaussians, are
the total number of stars in the cluster ($N_{c,max}$),
the number of field stars falling within the cluster
area ($N_{f,cri}$), and the cluster star density
profile ($\delta$). For each set of parameters we
have performed $100$ simulations and we have calculated
the average values of the studied quantities with their
corresponding standard deviations.

\section{Results from simulations}
\label{resultados}

\begin{figure*}
\centering
\includegraphics[width=17cm]{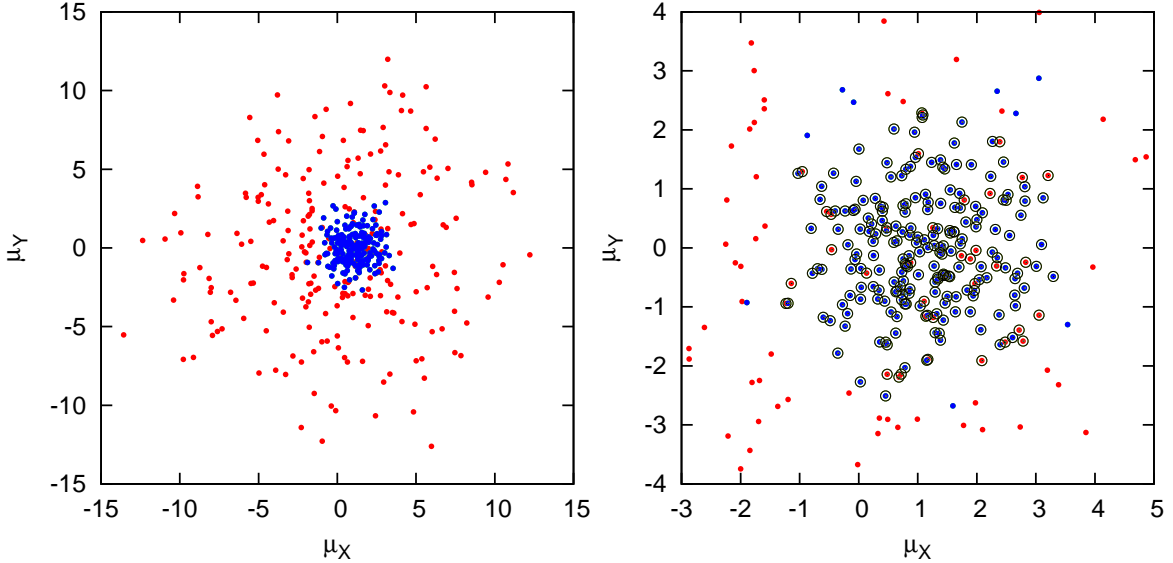}
\caption{Proper motion for the stars of a random simulation
with $N_{c,max}=N_{f,cri}=200$, $\delta=2$, and $R_s/R_c=1.1$
(see text for details of the meaning of each of these quantities).
Left panel shows the distribution for all the 442 simulated stars.
Red circles are the field stars centered at $(0,0)$ with $\sigma_f=5$
and blue circles are the $200$ cluster stars centered at $(1,0)$
with $\sigma_c=1$. Right panel is a magnification of the central
region in which we have marked with circles the stars whose
resulting cluster membership probabilities are higher than
$0.5$ according to the algorithm used.}
\label{movimientos}
\end{figure*}

For each simulation, we have calculated cluster membership
probabilities using the method described
in Section~\ref{membresias}. We have performed several
simulations varying the input parameters (the number of
stars in the cluster and in the field, centroid distance
in the proper motion space, and standard deviations) within
reasonable ranges. Except for minor differences, such as
that the error bars are higher when cluster and field
distributions are more similar, all the results and trends
remained essentially identical to those described in this
section.
Let us start showing how
the algorithm works. In Figure~\ref{movimientos} we
can see an example of a simulation
of a cluster of $200$ stars that it has
been adequately sampled with $R_s = 1.1 R_c$
The right panel
clearly shows the occasional but inevitable ``failures" of
the method. First, cluster stars falling in the tails of
their own distribution may not be recognized as members.
Second, field stars falling by chance below the cluster
distribution may be selected as probable members.

\begin{figure}
\resizebox{\hsize}{!}{\includegraphics{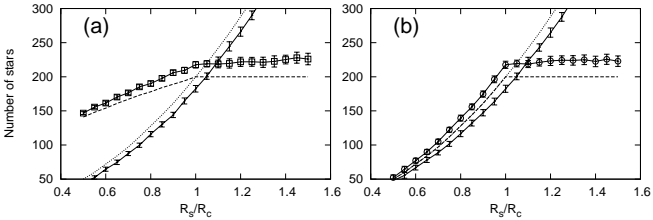}}
\caption{Calculated number of field and cluster stars
as a function of the sampling radius in units of the cluster
radius, $R_s/R_c$, for simulations with the same set of
parameters as Figure~\ref{movimientos}. (a) Simulation
with peaked density profile ($\delta=0.5$), assigned
members are indicated by squares connected by lines.
(b) Simulation with flat density profile ($\delta=2$),
members are indicated by circles connected by lines.
Assigned field stars are indicated by vertical bars
connected by lines, the length of the bars indicating
one standard deviation. The real numbers of simulated
stars are shown by dashed lines (cluster) and dotted
lines (field).}
\label{nstarsAB}
\end{figure}

\begin{figure}
\resizebox{\hsize}{!}{\includegraphics{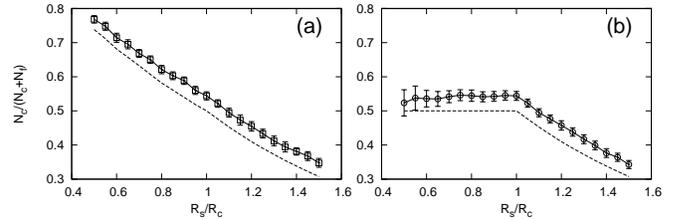}}
\caption{Calculated fraction of cluster stars as a
function of the sampling radius for the same simulations
as in Figure~\ref{nstarsAB}. The real (simulated) values are
shown by dashed lines.}
\label{fraccionAB}
\end{figure}

What would happen if we select a larger field? In order to
address this point we have calculated membership probabilities
as a function of the sampling radius. Here we are considering
as cluster members those stars having membership probabilities
$\geq 0.5$ in a Bayesian sense.
We have done several tests by using different
selection criteria and, as expected, the number of assigned
members depends on it, but the main results and trends
presented here remained unchanged. Figure~\ref{nstarsAB} shows
the number of stars classified as members
(we will denote it by $N_c$)
or non-members
($N_f$)
by the algorithm as a function of the sampling radius.
For these particular simulations the number of assigned
members $N_c$ is always higher than the real number of
cluster stars. Most of the cluster stars are well identified
but, as mentioned before, field stars falling below the
cluster distribution are also considered as members.
For the same reason the number of field stars is
always smaller than its expected value. For $R_s < R_c$
(subsampled cluster), $N_c$ increases with $R_s$ because
obviously the number of cluster stars in the sample
increases as $R_s$ increases. The rate at which this
occurs depends on the cluster density profile, that for
the simulations with $\delta=2$ in Figure~\ref{nstarsAB} is
exacly the same as for the field (homogeneous distribution).
For $R_s \geq R_c$ we observe a change in the behavior of
$N_c$. In this case we do not include new cluster stars
in the sample as $R_s$ increases, and $N_c$ increases
slightly because of the new field stars that erroneously
are classified as possible members. On the other hand,
field stars always increase at a rate roughly proportional
to $R_s^2$. It is easy to see that, in general, the fraction
of cluster stars (shown in Figure~\ref{fraccionAB}) should
be a decreasing function of $R_s$ for any cluster with
$\delta < 2$. Only for the extreme case of homogeneous
clusters the fraction of cluster stars remains constant
with $R_s$ for $R_s < R_c$.

\begin{figure}
\resizebox{\hsize}{!}{\includegraphics{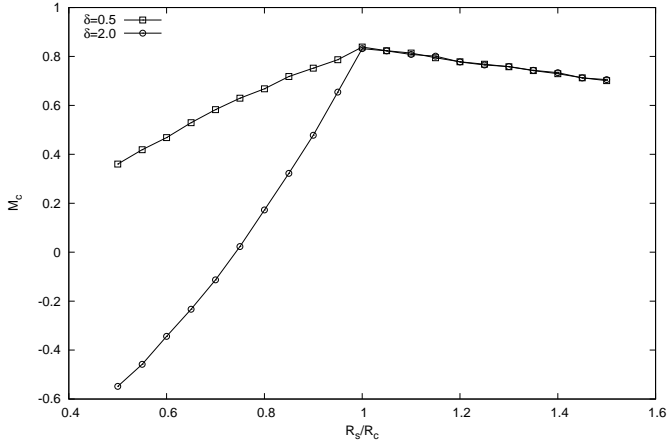}}
\caption{Matching fraction of the cluster (see text)
as a function of the sampling radius for the same simulations
as in Figure~\ref{nstarsAB}. The error bars are of the order of
the symbol sizes but they are not shown for clarity.}
\label{matching}
\end{figure}

Figures~\ref{nstarsAB} and \ref{fraccionAB} show the number of
stars classified as members, but we do not know whether this
classification is actually well done. In order to quantify
the correctness of the result we define the matching fraction
of the cluster $M_c$ as the net proportion of cluster stars
that are well classified. If $N_{ok}$ is the total number
of cluster stars correctly classified as members minus the number
of cluster stars incorrectly classified as non-members, then
$M_c=N_{ok}/N_{c,max}$. $M_c$ can be a negative number if the
number of misclassifications is higher than the number of
correct classifications and $M_c$ is exactly $1$ only when
the algorithm classifies correctly all the stars of the
cluster. In Figure~\ref{matching} we
see that the highest $M_c$ value occurs precisely when the
sampling radius equals the cluster radius. At smaller sampling
radii the matching fraction of the cluster obviously decreases
because the cluster is being subsampled. Interestingly, the
matching fraction is also smaller at $R_s > R_c$, but the
reason in this case is that more field stars are being
erroneously assigned to the cluster as $R_s$ increases.
The best classification is obtained when the sampling
radius is very close to the cluster radius although,
as expected, even in this case the matching fraction
does not reach its maximum value $M_c=1$.
However, the matching fraction is relatively high ($M_c=0.83$)
at $R_s=R_c$ and decreases slowly to $0.71$ at $R_s=1.5R_c$.
Moreover, the behaviors of $N_c$ and $N_f$ with $R_s$ are
very similar to the expected ones (Figures~\ref{nstarsAB} and
\ref{fraccionAB}). This is because both cluster and field
stars were simulated following perfect normal distributions
and, therefore, both populations can be well detected by
the algorithm
since it assumes the same kind of underlying
distribution.
When using real data the situation becomes
more complex, as discussed in the next section. 

\section{Results using real data}
\label{cumulos}

\begin{figure}
\resizebox{\hsize}{!}{\includegraphics{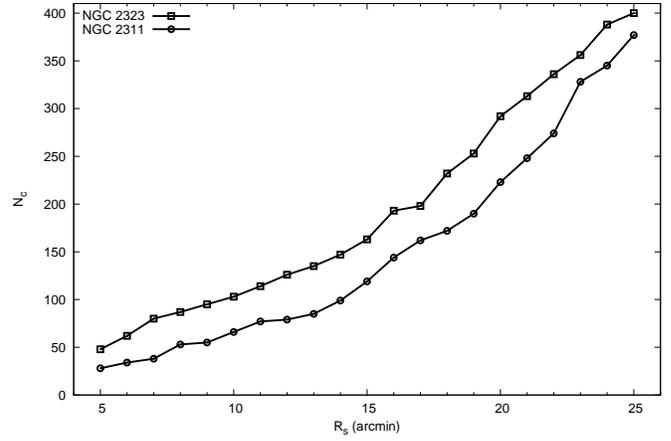}}
\caption{Number of cluster stars $N_c$ as a function of
the sampling radius $R_s$ in arcmin for the open clusters
NGC~2323 (squares connected by lines) and NGC~2311 (circles
connected by lines).}
\label{nstarsNGC}
\end{figure}

\begin{figure}
\resizebox{\hsize}{!}{\includegraphics{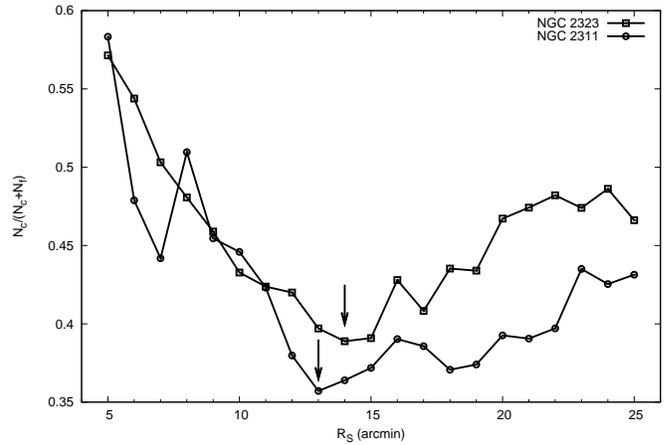}}
\caption{Fraction of cluster stars as a function of
the sampling radius for NGC~2323 (squares connected by
lines) and NGC~2311 (circles connected by lines).
Vertical arrows indicate the optimal
sampling radii (see text).}
\label{fraccionNGC}
\end{figure}

We use the CdC-SF Catalogue \citep{Vic09}, an astrometric
catalogue with a mean precision in the proper motions of
2.0 mas/yr (1.2 mas/yr for well measured stars,
typically $V < 14$). Given
the position of a known open cluster, we extract circular
fields of varying radius centred on it and then we calculate
membership probabilities by using the same algorithm as in
Section~\ref{resultados}. Here we analyze two open clusters
that are included in the area covered by
this catalogue: NGC~2323 (M~50) and NGC~2311.
In order to minimize even more the influence of
possible outliers on our results we further restrict the sample
to $|\mu | \leq 20$ mas/yr. The number of probable members $N_c$,
i.e. stars having membership probabilities higher than 0.5,
is shown in Figure~\ref{nstarsNGC} as
a function of the sampling radius. In general, $N_c$ always
increases with increasing $R_s$ and there are no relatively
flat regions analogous to those observed in Figure~\ref{nstarsAB}
for $R_s > R_c$. Without a previous knowledge of the approximate
value of the cluster radius, how can we determine which is the
most reliable result? This is not a trivial question given the
large uncertainties involved in the estimation or definition
of the cluster radius
(see discussion in Section~\ref{radios}). For
example, the radius of the total extent of NGC~2323 estimated
by different authors has been varying over the last years:
$10$ arcmin \citep{Cla98}, $16.7$ arcmin \citep{Nil02},
$15$ arcmin \citep{Kal03}, $22.2$ arcmin \citep{Kar05},
$17$ arcmin \citep[][using their own optical data]{Sha06} or
$22$ arcmin \citep[][using 2MASS data]{Sha06}.
Our calculations yield $N_c=198$ probable members in a field
of radius $R_s=17$ arcmin, but this number increases to
$N_c=336$ for $R_s=22$ arcmin. This means that there
could be more than $100$ undetected members if we use $R_s=17$
arcmin and the cluster radius is actually $R_c=22$ arcmin
or, on the contrary, more than $100$ spurious members if
we use $R_s=22$ arcmin and $R_c=17$ arcmin. The fraction
of cluster members is shown in Figure~\ref{fraccionNGC}.
The trend in which $N_c/(N_c+N_f)$ decreases with $R_s$
is qualitatively consistent with the expected behavior
(Figure~\ref{fraccionAB}). However, there is a $R_s$ value
from which the fraction of members increases as $R_s$
increases and, as mentioned in the previous section,
this behavior is possible only if $N_c$ increases
faster than $N_f$ does (i.e. at a rate higher than
$\sim R_s^2$). The only way this could happen is if
the algorithm is introducing many spurious members as
$R_s$ increases. In other words, there is a critical
$R_s$ value above which a significant number of spurious
members are erroneously included as part of the cluster
\citep[see also][]{Pia09}.
Here we call this critical value the optimal sampling
radius, $R_{s,opt}$, and obviously it is not recommended
to use a sampling radius larger than this value. From
Figure~\ref{fraccionNGC} we get $R_{s,opt} \simeq 14$
arcmin for NGC~2323 and $R_{s,opt} \simeq 13$ arcmin
for NGC~2311, but we have to point out that these
values are valid for the data we are using and,
in principle, they cannot be extrapolated to other
data sets.

\begin{figure}
\resizebox{\hsize}{!}{\includegraphics{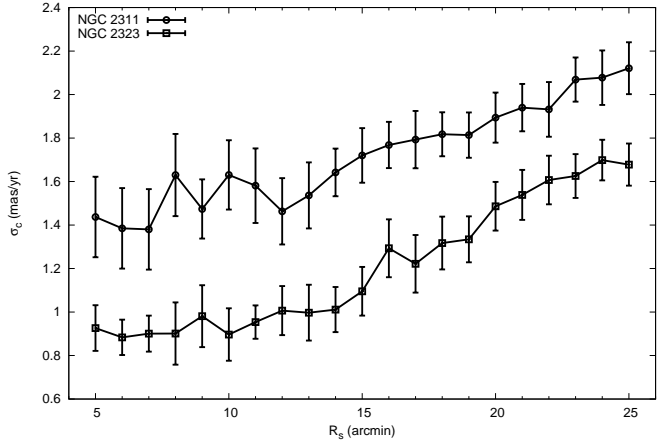}}
\caption{Estimated standard deviations as a function of
the sampling radius for the clusters NGC~2323 (squares
connected by lines) and NGC~2311 (circles connected by
lines). The bars indicate the uncertainties
obtained from bootstrapping.}
\label{sigmaNGC}
\end{figure}

The main reason behind the behavior observed in
Figure~\ref{fraccionNGC} lies in the disagreement
between the assumed and the ``true" underlying
distributions of proper motion of field stars.
A circular normal bivariate function is a good
representation of the cluster probability density
function (PDF), the standard deviation being the
result of observational errors that prevent the
intrinsic velocity dispersion of the cluster from
being completely resolved. However, it is known
that an elliptical normal bivariate function is
not always the best model for the field PDF
\citep[see discussions on this subject
in][]{Cab90,Uri94,Bal04,San09,Gri09}.
The combination of several factors, such as
galactic differential rotation or peculiar
motions, may affect the field star distribution
which usually tends to exhibit
non-gaussian tails.
Non-parametric models, which make no a
priori assumptions about the cluster or field star
distributions, have been introduced and used to
overcome this problem \citep[cf.][]{Cab90,Che97}.
It is interesting to note that both the classical
parametric and non-parametric methods agree
reasonably well with each other only for the
cases of nearly gaussian field distributions
\citep[see Figure~5 in][]{San09}.
When the number of field stars increases and the algorithm
tries to fit a gaussian function to the PDF, the fit
tends to produce a wider and flatter
function. As a consequence, the membership
probabilities (defined as the ratio of the cluster
to the total proper motion distribution function)
increases and therefore the number of assigned
members also increases. This effect is magnified
when the cluster distribution becomes ``contaminated"
by many field stars, because then the standard deviation
of the cluster tends to increases with the consequent
increasing of number of spurious members. The standard
deviations estimated for the two clusters under
consideration are shown in Figure~\ref{sigmaNGC}.
The error bars were estimated using bootstrap
techniques: the calculation is repeated on a series of 100
random resamplings of the data and the standard deviation
of the obtained set of values is taken as the associated
uncertainty.
The standard deviations remain nearly constant ($\sigma_c
\simeq 1.4-1.6$ for NGC~2311 and $\sigma_c \simeq 0.9-1.0$
for NGC~2323) in the region in which $R_s \lesssim R_{s,opt}$
(see also Figure~\ref{fraccionNGC}). This is the expected
behavior because, in principle, $\sigma_c$ should not depend
on the sample size. However, above the optimal sampling
radius we can see a gradual increase in $\sigma_c$ due
to the effect mentioned previously.

\subsection{Effectiveness of membership determination}

\begin{figure}
\resizebox{\hsize}{!}{\includegraphics{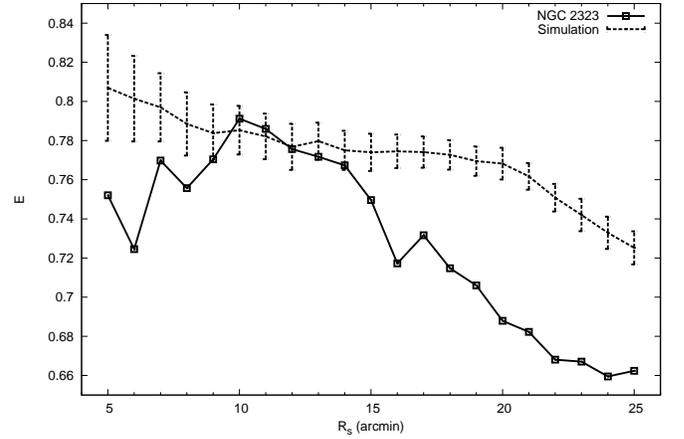}}
\caption{Effectiveness of membership determination (see
Equation~\ref{efec}) as a function of the sampling radius
for the open cluster NGC~2323 (open squares connected by
solid lines) and for simulations using parameter values
corresponding to those obtained for NGC~2323 (dashed
lines).}
\label{efecNGC}
\end{figure}

It is not possible in practice to quantify the degree of
correlation between identified and true cluster members,
such as the matching fraction in Figure~\ref{matching}.
Instead, we can use the concept of effectiveness of
membership determination which is set as \citep{Tia98,Wu02}
\begin{equation}
\label{efec}
E=1-
\frac{N \sum_{i=1}^N \{ p(i) \left[ 1-p(i) \right]
\}}{\sum_{i=1}^N p(i) \sum_{i=1}^N \left[ 1-p(i) \right]} \ \ ,
\end{equation}
where $p(i)$ is the membership probability of the $i$-th star
and $N$ is the sample size. This index measures how effective
the membership determination is in the sense of measuring
the separation between field and cluster populations in
the probability histogram. The higher the index $E$, the
more effective the membership determination. The maximum
$E$ value is obtained when there are two perfectly separated
populations of $N_c$ stars with membership probabilities
$p(i)=1$ and $N_f$ stars with $p(i)=0$. Figure~\ref{efecNGC}
shows $E$ for the open cluster NGC~2323 as a function
of the sampling radius. For the sake of comparison we
also show the result for simulations using the same
parameters as those for NGC~2323. Our most reliable
estimation for this cluster ($R_s=R_{s,opt}=14$ arcmin)
yielded the following values for the proper motions (in
mas/yr): 
$\mu_{x,c}=1.09$,
$\mu_{y,c}=1.13$,
$\sigma_{x,c}=\sigma_{y,c}=1.01$,
$\mu_{x,f}=+0.77$,
$\mu_{y,f}=-2.54$,
$\sigma_{x,f}=6.41$, and
$\sigma_{y,f}=5.84$.
According to the result shown in Figure~\ref{perfilNGC}
(next section),
we assume $R_c=20$ arcmin and $\delta=1.7$ for the cluster.
Additionally, we choose $N_{c,max}=250$ and $N_{f,cri}=500$
in order to get the measured values $N_c=147$ and $N_f=231$
at $R_s=14$ arcmin. The superimposed dashed lines in
Figure~\ref{efecNGC} are the average values (and their
standard deviations) for these simulations. The simulated $E$
value remains fairly constant (within the uncertainties) as
$R_s$ increases until the value $R_s \simeq R_c = 20$ arcmin,
beyond which it decreases at a relatively high rate. For
NGC~2323 we see that $E$ begins to decrease more rapidly as
$R_s$ increases just beyond $R_s \simeq R_{s,opt} = 14$ arcmin.
The best separations between cluster and field stars and
the agreement with the simulations are achieved in the
range $10 \lesssim R_s \lesssim 14$ arcmin.

\subsection{Cluster radius and optimal sampling radius}
\label{radios}

Basically what we are saying is that, at least
when using only kinematical criteria, the sample size
can substantially alter the results obtained
(the memberships and the rest of the properties
derived from there). Thus, the strategy of
choosing a field large enough to be sure of
covering more than the whole cluster has to be
taken with extreme caution, especially in dense
star fields. According to our simulations
(Section~\ref{resultados}), the best membership
estimation is achieved when $R_s \simeq R_c$.
This would seem an obvious result, given that
for $R_s < R_c$ the cluster is subsampled
whereas for $R_s > R_c$ the probability of
contamination by field stars is increased.
The important point here is: how well can
we know the cluster radius before estimating
memberships?
It is difficult to determine precisely the radius
of a cluster because the definition of radius is
ambiguous itself, since star clusters have no well
defined natural boundaries.
In this work we have used the usual
definition of $R_c$
as the radius of the circle containing
all the cluster members.
Most of the ``geometric" definitions
tend to overestimate the actual size,
especially for irregularly shaped
clusters \citep{Sch06}.
But this is not the main problem. The
problem is that the independent estimations
of cluster radii available in the literature
usually exhibit significant differences and
uncertainties. Angular sizes listed in
catalogues as
Webda\footnote{http://www.univie.ac.at/webda}
were compiled from
older references \citep[e.g.][]{Lyn87}
in which most of the apparent diameters
were estimated from visual inspection.
According to Webda $R_c = 7$ arcmin for
NGC~2323, but in the last years this
value has been triplicated \citep{Sha06}.
As mentioned above, it is an usual practice
to choose a field larger than the apparent
area covered by the cluster (taken from
the literature) for estimating membership
probabilities. But, at least when applying
the Sanders' method, assigned members will
spread throughout the whole selected area
because of the contamination by field
stars. It is probably not coincidental
that this is the case, for example,
for the probable members in the Dias
catalogue \citep{Dia02}. How reliable are
all the memberships that have been
derived from proper motions? It depends
on the ``real" $R_c$ values. Thus, again,
the ideal situation would be some kind
of robust estimation of the radius.

\begin{figure}
\resizebox{\hsize}{!}{\includegraphics{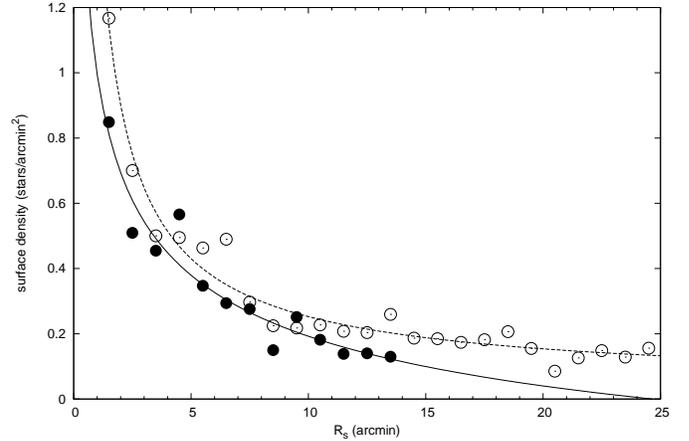}}
\caption{Radial density profiles for the cluster
NGC~2323 calculated for the cases $R_s=14$ arcmin
(solid circles) and $R_s=25$ arcmin (open circles).
Lines show the best fits for functions of the form
$\sim r^{\delta-2}$ (see Equation~\ref{densup}).
The solid line is for the case $R_s=14$ arcmin for
which $\delta \simeq 1.7$, and the dashed line
corresponds to $R_s=25$ arcmin for which $\delta
\simeq 1.2$.}
\label{perfilNGC}
\end{figure}

A commonly used procedure to determine (or define)
the cluster radius is based on the analysis of the
projected radial density profile. Usually, some
particular analytical function (for example, a
King-like model) is fitted to the density profile
and the cluster radius is extracted from this fit.
The last study dealing with a systematic determination
of cluster sizes based on objective and uniform
estimations of radial density profiles was done by
\citet{Kar04}. One limitation of this method is the
sensitivity of the fit to small variations in the
distribution of stars, especially for poorly
populated open cluster. The most reliable fits
are obtained using only the cluster members,
but then we are confronting again the problem
of membership determination. As an example,
let us consider Figure~\ref{perfilNGC} which
compares the density profiles obtained for the open
cluster NGC~2323 for two different sampling radii:
$R_s=R_{s,opt}=14$ arcmin and $R_s=25$ arcmin.
According to our results (section~\ref{cumulos})
our most reliable estimation is achieved when
$R_s=R_{s,opt}$. For this case, the best least
squares fit to a power law function suggests a
cluster radius in the range $\sim 20-25$ arcmin.
However, if we take a sample of size $R_s=25$
arcmin the contamination by field stars tends
to produce an overestimation of the star density
and both the index of the power law and the
estimated cluster radius change notoriously
(see Figure~\ref{perfilNGC}).
But the main drawback of this method is that
simple analytical fits are not always a good
representation for the stars distribution in
open clusters \citep{San09}.
The radius defined through a
fit to a density profile may be useful in
analyzing and comparing the properties of
several clusters systematically, but great
care must be taken when using these
model-dependent definition to estimate
the ``true" cluster radius. In fact,
the point where the fitted star density
equals the background (or drops to zero)
does not even necessarily agree with the
outer boundary of an open cluster.
In principle, new-born stars in a
young cluster spread out through the
region able to collapse gravitationally
to form them. At certain distance from
the high density peak in the molecular
cloud the required conditions are not
fulfilled anymore and the star formation
efficiency may decrease abruptly. So,
a radial star density distribution
which decays smoothly to $R_c$ may
not be always suitable,
especially for compact 
and/or very young star clusters.
Moreover, if the clusters exhibit
some degree of substructure this kind of
procedure yields totally unrealistic
results \citep{San09}.
Young embedded clusters often show
hierarchical structure \citep{Elm09}, so
that these methods cannot in principle be
applied to embedded clusters but only to
centrally concentrated open clusters.

Obviously, any reliable estimation of the
cluster radius ultimately depends on the
membership determination. 
Field star contamination may affect the
determination of $R_c$, and what we are showing
in this work is that this contamination can become a 
severe problem if it is not taken into consideration.
Furthermore, even though cluster and field populations
were well separated, the estimated radius would depend
on the limit magnitude if, for instance, there was mass
segregation. This kind of problems is particularly
relevant for the development of automated techniques
in which it is necessary to establish objective
criteria to decide the size of the sample to be
processed. What we are proposing here is to apply any
suggested method to several sample sizes $R_s$ and
analyse the behavior obtained.
It is difficult to give simple rules for evaluating
this behavior because the results will depend directly
on both the membership determination algorithm and the
input data. However, for the method considered in this
work, based on two underlying gaussian populations,
the basic procedure can be outlined as follows.
\begin{enumerate}
\item An upper limit for $R_s$ can be previously
estimated by fitting the spatial star density
to, for example, a King profile. The estimated tidal radius
(or, in order to be confident, twice its value) may be
considered an upper limit of the optimal sampling radius
and would define the range of $R_s$ values to be scanned.
\item For each $R_s$ value, cluster memberships and all
the relevant quantities (numbers of cluster stars and
field stars, centroids with their standard deviations,
effectiveness of membership determination) have to be
estimated.
\item The next step is to plot the number of cluster
members $N_c$ as a function of the sampling radius $R_s$.
If the membership determination works reasonably well,
meaning that it presents little contamination by field
stars, then we would observe a behavior as that seen in
Figure~\ref{nstarsAB}: $N_c$ increasing as $R_s$ increases
until some point (just when $R_s = R_c$) and then $N_c$
remaining approximately constant for higher $R_s$ values
(or increasing at a much slower rate). In this way, we
have a method to estimate the cluster size directly
from the data and the membership criteria without making
any additional assumptions. The optimal sampling radius
at which we get the best membership estimation is
precisely $R_{s,opt}=R_c$ (Fig.~\ref{matching})
\item If the parametric model does not adequately
describe the real data and/or if the internal noise
has not a simple structure, then the behavior of the
estimated parameters with $R_s$ would be different
from the expected one. If this were the case we
should plot the fraction of members $N_c/(N_c+N_f)$
versus $R_s$, where we should be able to identify the
optimal sampling radius $R_{s,opt}$ as the minimum in
this plot (Fig~\ref{fraccionNGC}). In absence of more
accurate information, this value would correspond to the
radius for which the membership classification is the
most reliable (with such a method in a given astrometric
catalogue).
\item Our experience indicates that the properties
derived from the Sanders' method tend to exhibit
some noise and it is not always easy to
identify the exact position of specific features (as
the minimum in the $N_c/(N_c+N_f)$ versus $R_s$ plot).
Some complementary strategies may be useful in
identifying or confirming the optimal sampling
radius. First, one can deal with the variation
of the proper motion standard deviation with radius.
The dispersion of the cluster proper motions should
display a change of slope at radius close to its
optimal value (Fig.~\ref{sigmaNGC}). Second, the
maximum of the effectiveness of membership
determination should also be around $R_{s,opt}$
(Fig.\ref{efecNGC}).
\end{enumerate}
The strategy proposed in this work, i.e. to
estimate and analyse cluster memberships as a
function of $R_s$, should in principle allow
for the identification of the optimal sampling
radius. However, we would like to emphasize
that it may not always be possible (or at least
not always unambiguous) to determine $R_{s,opt}$
in the way described above. For instance, for very
peaked cluster density profiles the change in $N_c$
at $R_s=R_c$ may be not pronounced enough for being
easily detected (e.g., Fig.~\ref{nstarsAB}a).
In spite of this, it still seems appropriate
and useful to perform this kind of tests before
any further analysis.

\section{Conclusions}
\label{conclusiones}

We have evaluated the performance of the commonly used
Sanders' method \citep{Vas58,San71,Cab85} in the
determination of star cluster memberships. In general,
the results depend on the radius of the field containing
the sampled cluster (the sampling radius,
$R_s$). The main reason for this
dependence lies in the differences between the
assumed gaussian and the true underlying proper
motion distributions. The contamination of cluster
members by field stars increases as the sampling radius
increases. The rate at which this effect occurs depends
on the intrinsic characteristics of the data set. There
is a threshold value of $R_s$ above which the identified
cluster members are highly contaminated by field stars and
the effectiveness of membership determination is relatively
small.
Thus, care must be taken when applying the Sanders'
method (just by itself or as part of a more extensive
procedure) especially when we do not have reliable
information about the real cluster radius and/or
when the sampling radius is larger the cluster
radius.
If this type of effects is not taken into
consideration in automated data analysis then significant
biases may arise in the derived cluster parameters.
The optimal sampling radius can be estimated by
plotting the number of cluster members and/or the fraction
of members as a function of the sampling radius. Moreover,
this type of analysis can also be used as an objective
procedure that can be applied systematically to determine
cluster radii.

\begin{acknowledgements}
We thank the referee for his/her comments which improved
this paper.
We acknowledge financial support from MICINN of Spain through
grant AYA2007-64052 and from Consejer\'{\i}a de Educaci\'on y
Ciencia (Junta de Andaluc\'{\i}a) through TIC-101 and TIC-4075.
N.S. is supported by a post-doctoral JAE-Doc (CSIC) contract.
E.J.A. acknowledges financial support from the Spanish MICINN
under the Consolider-Ingenio 2010 Program grant CSD2006-00070:
``First Science with the GTC".
\end{acknowledgements}

\end{document}